\documentclass[%
 reprint,
 amsmath,amssymb,
 aps,
]{revtex4-1}

\usepackage{graphicx}
\usepackage{dcolumn}
\usepackage{bm}
\pdfoutput=1 

\begin{document}

\preprint{APS/123-QED}

\title{Reaching the ideal glass transition by aging polymer films}

\author{Virginie M. Boucher$^1$}

\author{Daniele Cangialosi$^1$}%
 \email{Second.Author@institution.edu}
 \author{Angel Alegr\'{\i}a$^{1,2}$}
 \author{Juan Colmenero$^{1,2,3}$}

\affiliation{%
 $^1$Centro de F\'{\i}sica de Materiales, Paseo Manuel de Lardizabal 5, 20018 San Sebasti\'{a}n, Spain}%

 \homepage{http://www.Second.institution.edu/~Charlie.Author}
\affiliation{
 $^2$Departamento de F\'{\i}sica de Materiales (UPV/EHU), Apartado 1072, 20080 San Sebasti\'{a}n, Spain\\
}%
\affiliation{
 $^3$Donostia International Physics Center, Paseo Manuel de Lardizabal 4, 20018 San Sebasti\'{a}n, Spain
}%

\date{\today}

\begin{abstract}
Searching for the ideal glass transition, we exploit the ability of glassy polymer films to explore low energy states in remarkably short time scales. We use 30 nm thick polystyrene (PS) films, which in the supercooled state basically display the bulk polymer equilibrium thermodynamics and dynamics. We show that in the glassy state, this system exhibits two mechanisms of equilibrium recovery. The faster one, active well below the kinetic glass transition temperature ($T_g$), allows massive enthalpy recovery. This implies that the 'fictive' temperature ($T_f$) reaches values as low as the predicted Kauzmann temperature ($T_K$) for PS. Once the thermodynamic state corresponding to $T_f = T_K$ is reached, no further decrease of enthalpy is observed. This is interpreted as a signature of the ideal glass transition. 
\begin{description}

\item[PACS numbers]
64.70.pj, 65.60.+a.

\end{description}
\end{abstract}

\pacs{64.70.pj, 65.60.+a}
\maketitle


Glasses are a special class of materials characterized by the disordered structure of liquids and the mechanical properties of solids. The most common route to form a glass is by cooling a liquid through its melting temperature avoiding crystallization. In this case, before forming a glass, the supercooled state is explored, that is, a system in metastable equilibrium \cite{Debenedetti,Schmelzer}. A problem of extraordinary fundamental importance arises from the observation that the supercooled state exhibits second order thermodynamic properties typical of a liquid, in particular, larger than those of the crystal. Given the difference between the supercooled and crystal specific heats, Kauzmann noticed that there will be a temperature ($T_K$) at which the entropy of the supercooled liquid and that of the crystal would be equal \cite{Kauzmann1948}. Gibbs and Di Marzio \cite{GibbsDiMarzio} subsequently theorized that a second order thermodynamic transition to an ideal glass would occur. This transition would avoid that the paradoxical scenario of a disordered glass with entropy smaller than that of the crystal is realized. Investigating the intriguing scenario of a liquid with the entropy of a crystal is prevented by the occurrence of the so-called kinetic glass transition \cite{Tropin2015} transforming the supercooled liquid into a glass. Once formed, non-equilibrium glasses evolve towards the metastable equilibrium of the supercooled liquid by decreasing their energy, a well-known phenomenon addressed as physical aging or structural recovery \cite{Struik1977}.

Motivated by the search for the second order thermodynamic transition (the ideal glass transition), different routes to low energy glassy states have been recently explored. One of them is the investigation of fossil glassy amber naturally aged for millions of years \cite{McKenna2013}, showing fictive temperature ($T_f$) $-$ i.e. the temperature corresponding to the intercept of the extrapolated glass and equilibrium lines for a given glass (see Figure 1) $-$ significantly lower than $T_g$, the glass transition temperature.  Another route, much faster, is the preparation of vapor deposited glasses which, when obtained in appropriate experimental conditions, exhibit $T_f$ values as low as about 30-40 K below $T_g$ for organic glass formers \cite{Ediger2007}. However, in both cases, the lowest $T_f$ reached is still far from $T_K$.

\begin{figure} [htb!]
\centering
\includegraphics[width=.5\textwidth, angle=0]{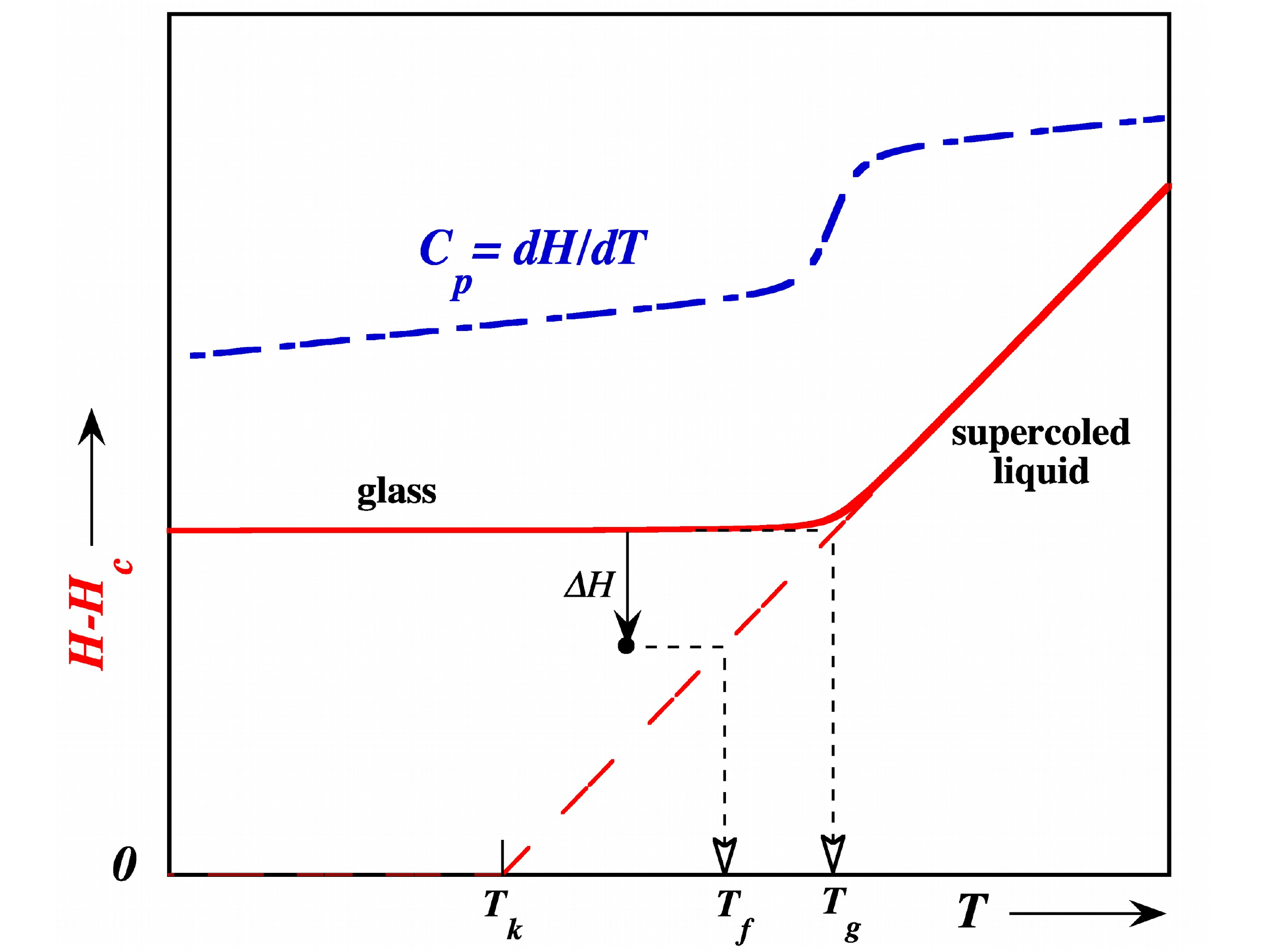}
\caption{Schematic representation of the temperature dependence of the specific heat, the enthalpy and the excess entropy of a glass forming liquid cooled down at a given rate. $T_g$, $T_K$ and the way $T_f$ is connected to the enthalpy recovered ($\Delta H$) during isothermal aging are shown.}
\label{Fig1}
\end{figure}

In recent years, several studies showed that glasses where one or more dimensions are restricted to the nanometer scale exhibit depressed $T_g$ in comparison to the corresponding bulk glass formers. This is especially pronounced in freestanding polystyrene (PS) films \cite{Forrest1996}. These can exhibit $T_g$ values of several tens of Kelvin lower than bulk for film thicknesses of the order of 30 nm.  Importantly, despite the $T_g$ depression, 
 recent investigations indicate that: $i$) the linear dynamics, that is, the rate of spontaneous fluctuations, of freestanding PS films is bulk-like, with the exception of a surface layer of a few nanometers \cite{Ediger2011}; $ii$) the thermodynamics of melt PS, and in general different kind of glass formers, in metastable equilibrium is essentially thickness invariant down to $\sim$ 30 nm \cite{Baschnagel2006,Douglas2012,Vilgis2014,Lipson2015}. For such thickness, a modest excess specific volume as compared to the bulk is observed and the coefficient of thermal expansion is bulk-like \cite{Lipson2015,Kanaya2011}. In particular a density difference of about 0.5 $\%$  is observed, implying a $T_K$ difference of less than 5 K \cite{Lipson2015}. Furthermore, molecular dynamics simulations in model systems of polymeric glass-formers indicate that perturbations of the bulk thermodynamics is limited to a thin layer close to the free \cite{Baschnagel2006,Douglas2012} or weakly interacting \cite{Vilgis2014} interface. Finally, perturbations of the melting temperature \cite{Frank2003,Napolitano2015} and heat of fusion \cite{McKenna2005} are generally observed only for confinement length scales shorter than 30 nm.
 
 Altogether, these findings indicate that glasses with large "free" interface \cite{Napolitano2012} exhibit decoupling between the rate of spontaneous fluctuations, exhibiting essentially bulk-like behavior, and the rate of non-linear recovery of equilibrium (for discussion on these aspects see the following reviews \cite{Cangialosi2015,Kremer2015,Priestley2015}). This fact is especially evident in the glassy state, where equilibrium is achieved considerably faster than in bulk \cite{Boucher2012}.      

Here, we exploit the ability of freestanding films to recover equilibrium within relatively short time scales to explore thermodynamic states with low energies. Furthermore, we take advantage of the presence of two mechanisms of structural recovery, recently found in experiments \cite{Cangialosi2013,Roth2011} (see Supplemental Material for details) and rationalized within the random first order transition (RFOT) theory \cite{Wolynes2014}. In doing so, we show that, for 30 nm thick PS films, the "fast" mechanism of equilibration, showing up below about 300 K, allows considerable enthalpy recovery and thereby a significant reduction of $T_f$. Notably this occurs in time scales not longer than several days. When aged in the low temperature range, these films are able to explore the landscape at energies corresponding to $T_f$ values of the order of the $T_K$ calculated for bulk PS ($\sim$ 278-280 K). After prolonged aging at lower temperatures, no further decrease of $T_f$ is observed. This indicates that the lowest energy state of the film is achieved and, thereby, the thermodynamic (ideal) glass transition exists. 


High molecular weight polystyrene (PS) ($M_w$ = 1408 Kg/mol and $M_w$/$M_n$ = 1.17) from Polymer Source Inc. was employed to produce films. After dissolving in toluene, the so-obtained solution was spin coated on a glass substrate. The films were subsequently dried at 433 K for several days to remove any trace of solvent and release the stress induced by spin-coating \cite{Reiter2005,Kremer2008}. The final films thickness was about 30 nm as revealed by atomic force microscopy (AFM). Films were finally floated on water and removed from the substrate. Several hundreds films were stacked in a pan for differential scanning calorimetry (DSC) to achieve a sample mass of about 4 mg. 

Specific heat measurements were done on a DSC-Q2000 calorimeter from TA-Instruments, calibrated with melting indium. All measurements were carried out in nitrogen atmosphere. To prevent stacked PS films to interdiffuse the maximum temperature employed never exceeded 393 K and the time spent at such temperature was always shorter than several minutes. As estimated in Ref. \cite{Wool1991}, for the very high molecular weight PS employed in the present study, this ensures to avoid interdiffusion of the layers. The absence of interdiffusion of the stacked PS layers was also verified by the reproducibility of the $T_g$ measurements on the same sample after several temperature cycles. 

For the study of enthalpy recovery, the samples underwent a heating ramp to a temperature of 393 K, that is, above the films' $T_g$ (348 K on cooling at 20 K/min \cite{Cangialosi2012}), with a stabilization period of 1 min, to erase the materials previous thermal history. They were subsequently cooled down at 20 K/min to 233 K before stabilizing at the temperature used for structural recovery, $T_a$. Then, they were aged in the calorimeter for times from minutes to several days before being cooled down to 183 K at a cooling rate of 20 K/min, prior to reheating at 10 K/min to 393 K for data collection.

The amount of enthalpy relaxed during aging of a glass for a period of time $t_a$ at a given temperature $T_a$ was evaluated by integration of the difference between the thermograms of the aged and unaged samples subsequently recorded, according to the relation:

\begin{equation}
\centering
\Delta H(T_a,t_a)=\int^{T >> T_g}_{T << T_g} (C^a_p(T)-C^u_p(T))\,dT
\label{Eq1}
\end{equation}
 				
In this equation, $C^a_p$(T) and $C^u_p$(T) are the heat capacity measured after the annealing and that of the unannealed sample, respectively.

The fictive temperature ($T_f$), that is the temperature at which a glass with given thermo-mechanical history would be at equilibrium, was determined from heating scans after given aging time and temperature using the Moynihan method \cite{Moynihan1976}:

\begin{equation}
\centering
\int_{T_f}^{T >> T_g} (C_{pm} - C_{pg}) dT = \int_{T << T_g}^{T >> T_g} (C_{p} - C_{pg}) dT
\label{Eq2}
\end{equation}

where $C_{pm}$, $C_{pg}$ and $C_p$ are the melt, glass and experimental specific heats. 

\begin{figure} [htb!]
\centering
\includegraphics[width=.35\textwidth, angle=270]{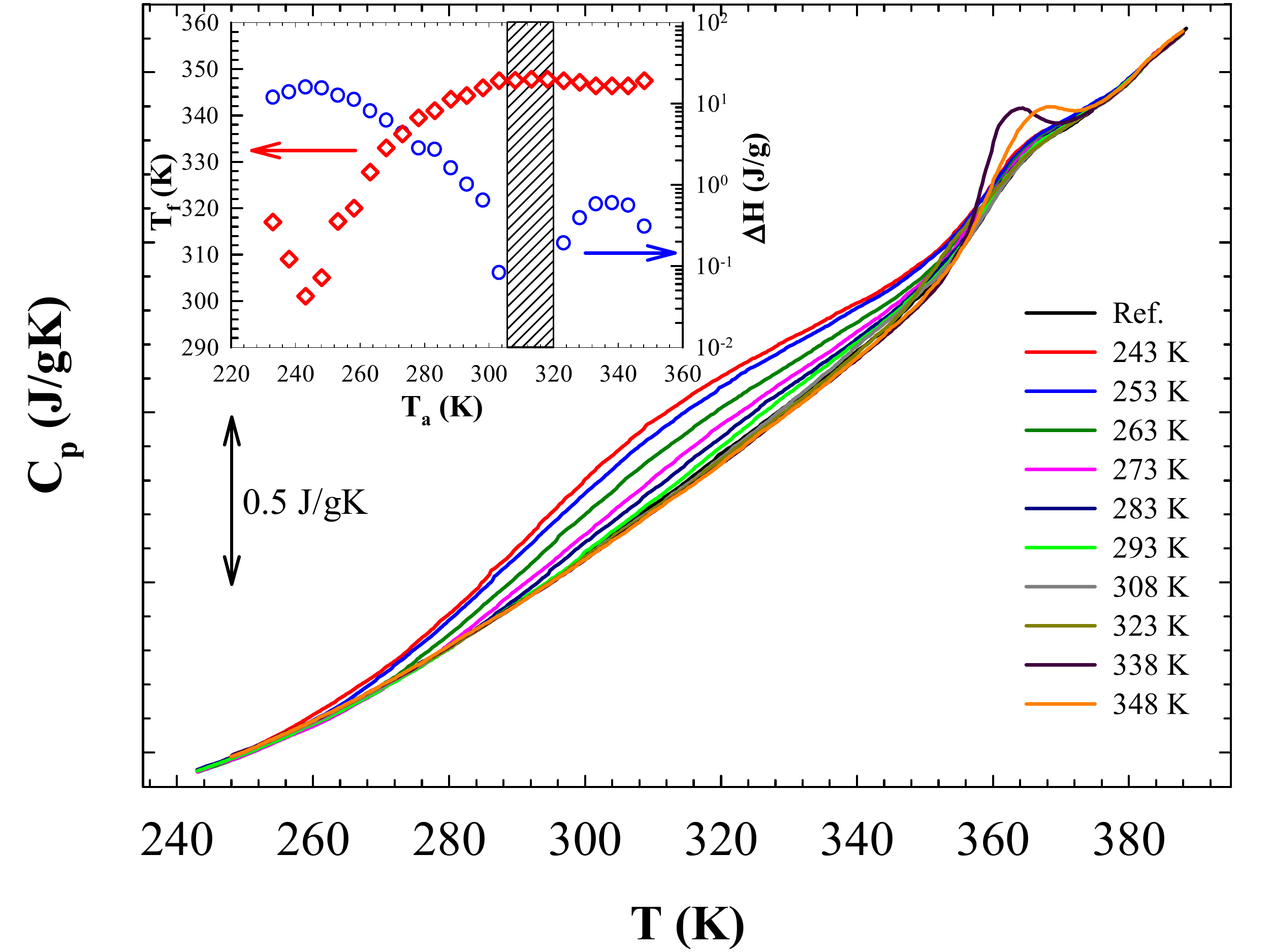}
\caption{Heat capacity as a function of temperature for 30 nm PS films after 480 min aging at the indicated aging temperatures. The reference curve corresponds to a sample heated up just after cooling. The inset shows the resulting recovered enthalpy (right axis) and $T_f$ (left axis) as a function of the aging temperature. The hatched area indicates the temperature range where negligible enthalpy recovery is observed in 480 min aging time.}
\label{Fig2}
\end{figure}

Figure \ref{Fig2} shows DSC heating scans for samples aged over 480 min at the indicated aging temperatures ($T_a$). Samples aged at $T$ $>$ 323 K exhibit the standard behavior, consisting of an overshoot at temperatures in proximity of the films' $T_g$. Samples aged between 308 and 323 K show negligible decrease in enthalpy. This is apparent from the superposition of specific heat curves. However, when the aging temperature is further decreased, a broad pronounced overshoot appears in a temperature range between $\sim$ 270 and 350 K. The presence of such overshoot implies the presence of a "fast" recovery mechanism, active at low temperatures, in line with previous experiments \cite{Roth2011}.

The inset of Figure \ref{Fig2} shows the amount of recovered enthalpy obtained from the DSC scans after 480 min aging and the corresponding values of $T_f$ as a function of the aging temperature. Both magnitudes exhibit non-monotonous temperature behavior. In particular, the recovered enthalpy shows a clear minimum and, as a consequence, $T_f$ goes through a maximum. Importantly, below $\sim$ 300 K, $T_f$ decreases drastically and, after 480 min aging at 243 and 248 K, is about 40 K lower than the film $T_g$. At lower temperatures a reduction of the recovered enthalpy and, thereby, an increase of $T_f$ is observed.

\begin{figure} [htb!]
\centering
\includegraphics[width=.35\textwidth, angle=270]{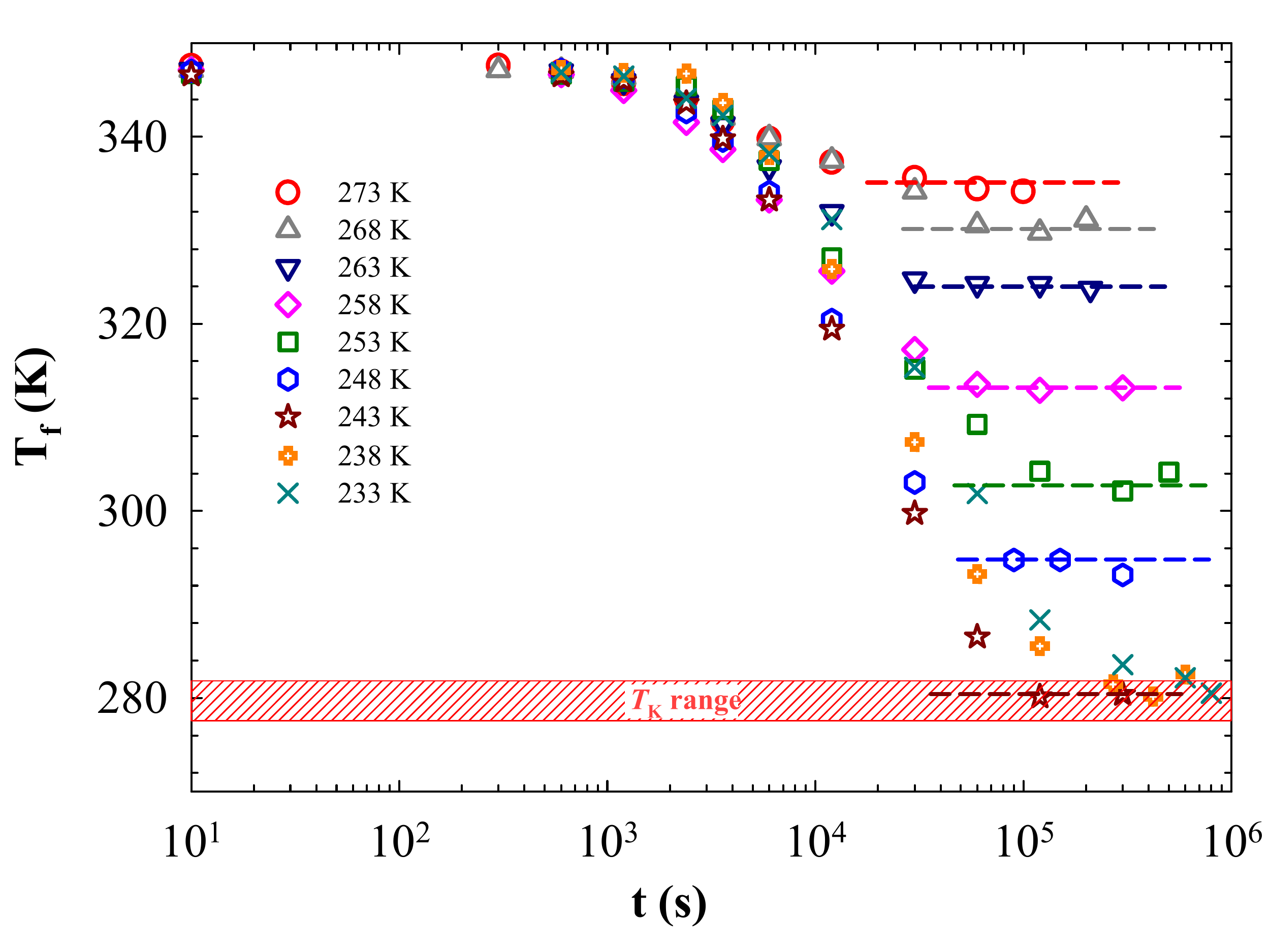}
\caption{Fictive temperature $T_f$ as a function of aging time at the indicated aging temperatures. The hatched area shows the range of variation of $T_K$ according to previous estimations \cite{Miller1970,Hodge1999,Cangialosi2005}. The dashed lines are guides for the eye to indicate the $T_f$ reached after prolonged aging.}
\label{Fig3}
\end{figure}

Enthalpy recovery experiments were carried out over a wide range of aging times and temperatures below 300 K (see Supplemental Material). A summary of the aging time dependence of $T_f$ is presented in Figure \ref{Fig3}. For aging temperatures above 243 K, the $T_f$ values at the plateau after long aging times decreases with $T_a$. A tremendous decrease of $T_f$ is observed at the lowest investigated $T_a$. Remarkably, the lowest $T_f$, obtained at $T_a$ $\leq$ 243 K, is $\sim$ 278-280 K. This is $\sim$ 70 K smaller than the film $T_g$. Such a small $T_f$ is achieved after less than two days aging. 
This is indicative of the low activation energy of the fast mechanism of equilibration, as already found in bulk PS \cite{Cangialosi2013}. At aging temperatures lower than 243 K, the $T_f$ values at the long aging time plateau do not experience further decrease, remaining at $\sim$ 278-280 K.

\begin{figure} [htb!]
\centering
\includegraphics[width=.45\textwidth, angle=0]{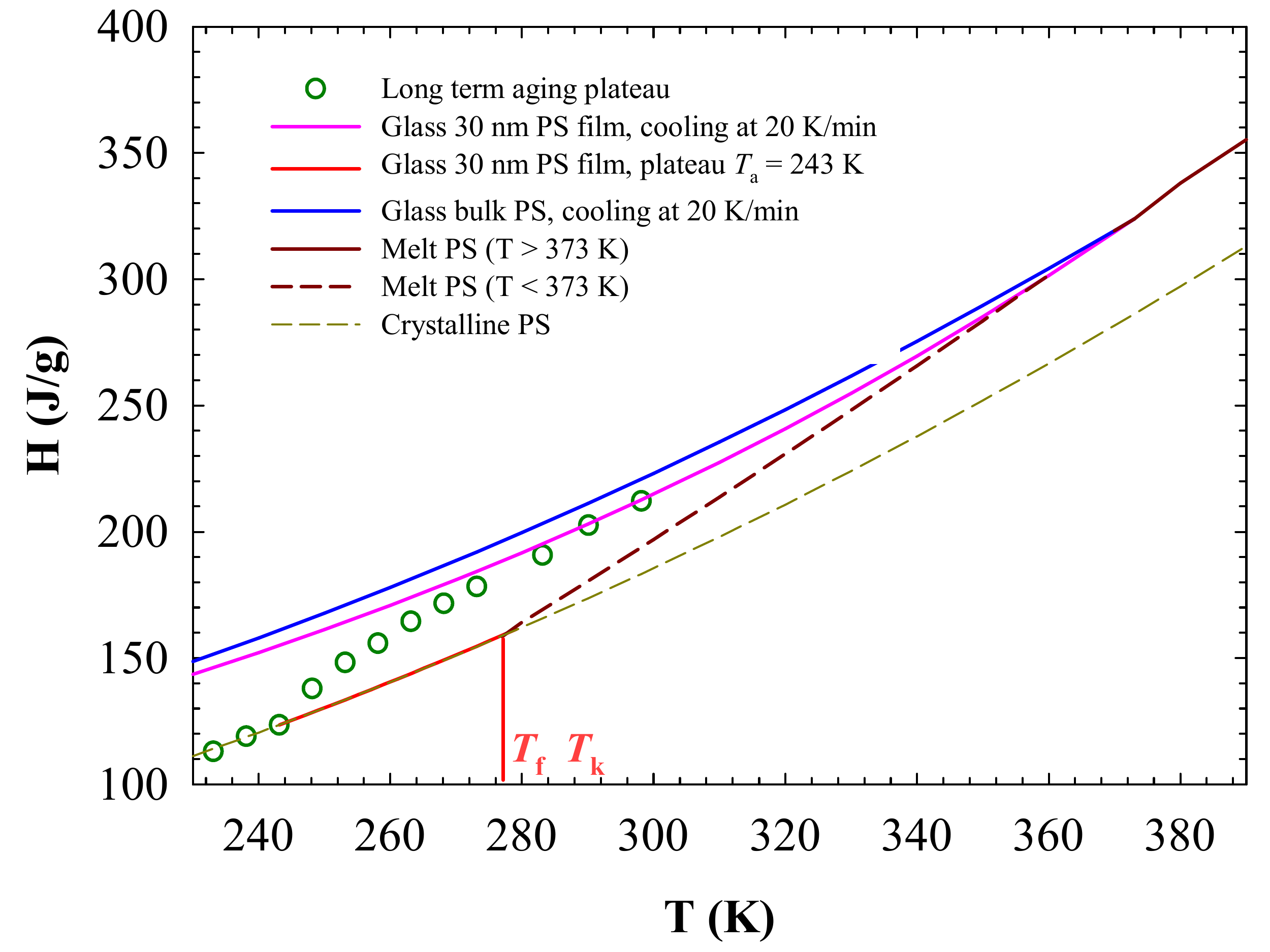}
\caption{Enthalpy as a function of temperature in PS glasses obtained in the indicated conditions; in the melt state; and after aging to a plateau at different aging temperatures below about 300 K. The thermodynamic state after aging at $T_a$ $\leq$ 243 K to a plateau with $T_f$ $\sim$ $T_K$ is also shown (red line).}
\label{Fig4}
\end{figure}

Figure \ref{Fig4} shows the enthalpy, obtained from DSC data, as a function of temperature for 30 nm PS films after different thermal histories. In particular, the enthalpy corresponding to the glass cooled down at 20 K/min and that obtained after prolonged aging below $\sim$ 300 K are shown. The enthalpy of the supercooled state and its extrapolation to temperatures below $T_g$ and that of bulk glassy PS cooled at 20 K/min are also shown. Such bulk properties are taken from the ATHAS databank \cite{ATHAS}. The thermodynamic diagram of Figure \ref{Fig4} indicates that the enthalpy reached after long time aging exhibits a kink at about 243 K.

Thermodynamic data from the ATHAS databank also allow determining $T_K$ by extrapolating the melt entropy to temperatures lower than $T_g$ (see Note \footnote{Data for crystalline PS, though not experimentally available due to the inability of this polymer to crystallize, are either obtained theoretically or relying on the ability of isotactic PS to crystallize. In this case, as shown by O'Reilly and coworkers \cite{OReilly1965}, the lack of effect of tacticity on equilibrium thermodynamic properties is exploited.} for details). This gives $T_K$ $\sim$ 280 K, which nicely matches with previous determinations \cite{Miller1970,Hodge1999,Cangialosi2005}. The range of variation of the reported values of $T_K$ is shown in Figure \ref{Fig3} (hatched area). 


Most importantly, data points of Figure \ref{Fig4} and the long aging time limit of $T_f$ shown in Figure \ref{Fig3} indicate that the lowest enthalpy state achieved corresponds to $T_f$ $\sim$ $T_K$. Provided that the $T_f$ completely defines the thermodynamic state of the system, the equality between $T_f$ and $T_K$ carries as a natural consequence that the entropy of the glass achieves a value equal to the crystalline counterpart. This, together with the kink of the enthalpy shown in Figure \ref{Fig4}, can be interpreted as a signature of the occurrence of a second order thermodynamic transition. In the kinetic conditions employed in the present study, that is, cooling at 20 K/min and isothermal aging at different temperatures, the transformation from one phase to the other occurs at 243 K.  

The observed transformation to a thermodynamic state with vanishing excess entropy can be considered as a signature of the existence of the ideal glass transition, long ago theorized by Gibbs and Di Marzio \cite{GibbsDiMarzio} and still subject of intense debate \cite{Debenedetti2013}. The existence of the ideal glass transition has been predicted by several theoretical approaches, such as the RFOT theory \cite{Wolynes1989} and the replica formulation of Mezard and Parisi \cite{Parisi1999}. Recently, molecular dynamics simulations on an equilibrated randomly pinned Lennard-Jones glass-former show the existence of the ideal glass transition \cite{Cammarota2012,Kob2015}, in agreement with our experimental results. However, the finite aging time needed to reach $T_f$ $\sim$ $T_K$ implies that there exists at least one non-diverging relaxation time at temperatures smaller than $T_K$. This has also been suggested by several experimental studies \cite{Dyre2008,McKenna2013,Scopigno2015}. Importantly, molecular dynamic simulations showed that the presence of a molecular mechanism of relaxation with finite time scale \cite{Chakrabarty2015} can be detected on the ideal glass obtained by random pinning \cite{Cammarota2012,Kob2015}. This does not rule out the possibility that a diverging time scale exists in the ideal glass at $T_K$. The RFOT theory \cite{Wolynes1989} predicts that glasses approaching the ideal state actually show two molecular mechanisms of relaxation; one exhibiting diverging time scale at $T_K$ and a non-diverging mechanism with relatively low activation energy \cite{Wolynes2014}.

In conclusion, investigating the enthalpy recovery of 30 nm thick PS films, we show the presence of two mechanisms of equilibration, similarly to bulk PS \cite{Cangialosi2013}. Given its low activation energy and its short equilibration time scale in 30 nm thick films, the "fast" mechanism is able to draw the glass deep in the energy landscape. This results in $T_f$ values, after aging times of the orders of several days, as low as the $T_K$ of PS. In a bulk glass, this would require aging times likely longer than millions of years \cite{McKenna2013}. Once $T_f$ $\sim$ $T_K$ is reached, no further decrease of $T_f$ is observed. Hence, the existence of the thermodynamic (ideal) glass transition is experimentally verified for the first time. This result has deep implications in the comprehension of the glass transition: $i)$ it provides clear evidence of its thermodynamic nature; $ii)$ it shows that a disordered liquid can achieve entropy equal to that of the ordered crystalline solid.

\section*{Acknowledgements}

The authors acknowledge the University of the Basque Country and Basque Country Government (Ref. No. IT-654-13 (GV)), Depto. Educaci\'{o}n, Universidades e investigaci\'{o}n; and Spanish Government (Grant No. MAT2015-63704-P, (MINECO/FEDER, UE)) for their financial support.


\begin{thebibliography}{45}%
\makeatletter
\providecommand \@ifxundefined [1]{%
 \@ifx{#1\undefined}
}%
\providecommand \@ifnum [1]{%
 \ifnum #1\expandafter \@firstoftwo
 \else \expandafter \@secondoftwo
 \fi
}%
\providecommand \@ifx [1]{%
 \ifx #1\expandafter \@firstoftwo
 \else \expandafter \@secondoftwo
 \fi
}%
\providecommand \natexlab [1]{#1}%
\providecommand \enquote  [1]{``#1''}%
\providecommand \bibnamefont  [1]{#1}%
\providecommand \bibfnamefont [1]{#1}%
\providecommand \citenamefont [1]{#1}%
\providecommand \href@noop [0]{\@secondoftwo}%
\providecommand \href [0]{\begingroup \@sanitize@url \@href}%
\providecommand \@href[1]{\@@startlink{#1}\@@href}%
\providecommand \@@href[1]{\endgroup#1\@@endlink}%
\providecommand \@sanitize@url [0]{\catcode `\\12\catcode `\$12\catcode
  `\&12\catcode `\#12\catcode `\^12\catcode `\_12\catcode `\%12\relax}%
\providecommand \@@startlink[1]{}%
\providecommand \@@endlink[0]{}%
\providecommand \url  [0]{\begingroup\@sanitize@url \@url }%
\providecommand \@url [1]{\endgroup\@href {#1}{\urlprefix }}%
\providecommand \urlprefix  [0]{URL }%
\providecommand \Eprint [0]{\href }%
\providecommand \doibase [0]{http://dx.doi.org/}%
\providecommand \selectlanguage [0]{\@gobble}%
\providecommand \bibinfo  [0]{\@secondoftwo}%
\providecommand \bibfield  [0]{\@secondoftwo}%
\providecommand \translation [1]{[#1]}%
\providecommand \BibitemOpen [0]{}%
\providecommand \bibitemStop [0]{}%
\providecommand \bibitemNoStop [0]{.\EOS\space}%
\providecommand \EOS [0]{\spacefactor3000\relax}%
\providecommand \BibitemShut  [1]{\csname bibitem#1\endcsname}%
\let\auto@bib@innerbib\@empty
\bibitem [{\citenamefont {Debenedetti}(1996)}]{Debenedetti}%
  \BibitemOpen
  \bibfield  {author} {\bibinfo {author} {\bibfnamefont {P.~G.}\ \bibnamefont
  {Debenedetti}},\ }\href@noop {} {\emph {\bibinfo {title} {Metastable Liquids:
  Concepts and Principles}}}\ (\bibinfo  {publisher} {Princeton University
  Press, Princeton},\ \bibinfo {year} {1996})\BibitemShut {NoStop}%
\bibitem [{\citenamefont {Schmelzer}\ and\ \citenamefont
  {Gutzow}(2011)}]{Schmelzer}%
  \BibitemOpen
  \bibfield  {author} {\bibinfo {author} {\bibfnamefont {J.~W.~P.}\
  \bibnamefont {Schmelzer}}\ and\ \bibinfo {author} {\bibfnamefont {I.~S.}\
  \bibnamefont {Gutzow}},\ }\href@noop {} {\emph {\bibinfo {title} {Glasses and
  the glass transition}}}\ (\bibinfo  {publisher} {Wiley-VCH, Weinheim},\
  \bibinfo {year} {2011})\BibitemShut {NoStop}%
\bibitem [{\citenamefont {Kauzmann}(1948)}]{Kauzmann1948}%
  \BibitemOpen
  \bibfield  {author} {\bibinfo {author} {\bibfnamefont {W.}~\bibnamefont
  {Kauzmann}},\ }\href@noop {} {\bibfield  {journal} {\bibinfo  {journal}
  {Chem. Rev.}\ }\textbf {\bibinfo {volume} {43}},\ \bibinfo {pages} {219}
  (\bibinfo {year} {1948})}\BibitemShut {NoStop}%
\bibitem [{\citenamefont {Gibbs}\ and\ \citenamefont
  {DiMarzio}(1958)}]{GibbsDiMarzio}%
  \BibitemOpen
  \bibfield  {author} {\bibinfo {author} {\bibfnamefont {J.~H.}\ \bibnamefont
  {Gibbs}}\ and\ \bibinfo {author} {\bibfnamefont {E.~A.}\ \bibnamefont
  {DiMarzio}},\ }\href@noop {} {\bibfield  {journal} {\bibinfo  {journal} {J.
  Chem. Phys.}\ }\textbf {\bibinfo {volume} {28}},\ \bibinfo {pages} {373}
  (\bibinfo {year} {1958})}\BibitemShut {NoStop}%
\bibitem [{\citenamefont {Tropin}\ \emph {et~al.}(2016)\citenamefont {Tropin},
  \citenamefont {Schmelzer},\ and\ \citenamefont {Aksenov}}]{Tropin2015}%
  \BibitemOpen
  \bibfield  {author} {\bibinfo {author} {\bibfnamefont {T.~V.}\ \bibnamefont
  {Tropin}}, \bibinfo {author} {\bibfnamefont {J.~W.}\ \bibnamefont
  {Schmelzer}}, \ and\ \bibinfo {author} {\bibfnamefont {V.~L.}\ \bibnamefont
  {Aksenov}},\ }\href@noop {} {\bibfield  {journal} {\bibinfo  {journal} {Phys.
  Usp.}\ }\textbf {\bibinfo {volume} {59}},\ \bibinfo {pages} {42} (\bibinfo
  {year} {2016})}\BibitemShut {NoStop}%
\bibitem [{\citenamefont {Struik}(1977)}]{Struik1977}%
  \BibitemOpen
  \bibfield  {author} {\bibinfo {author} {\bibfnamefont {L.~C.~E.}\
  \bibnamefont {Struik}},\ }\href
  {http://books.google.es/books?id=Yc8EPwAACAAJ} {\emph {\bibinfo {title}
  {Physical aging in amorphous polymers and other materials}}}\ (\bibinfo
  {publisher} {Technische Hogeschool Delft.},\ \bibinfo {year}
  {1977})\BibitemShut {NoStop}%
\bibitem [{\citenamefont {Zhao}\ \emph {et~al.}(2013)\citenamefont {Zhao},
  \citenamefont {Simon},\ and\ \citenamefont {McKenna}}]{McKenna2013}%
  \BibitemOpen
  \bibfield  {author} {\bibinfo {author} {\bibfnamefont {J.}~\bibnamefont
  {Zhao}}, \bibinfo {author} {\bibfnamefont {S.~L.}\ \bibnamefont {Simon}}, \
  and\ \bibinfo {author} {\bibfnamefont {G.~B.}\ \bibnamefont {McKenna}},\
  }\href@noop {} {\bibfield  {journal} {\bibinfo  {journal} {Nat. Commun.}\
  }\textbf {\bibinfo {volume} {4}},\ \bibinfo {pages} {1783} (\bibinfo {year}
  {2013})}\BibitemShut {NoStop}%
\bibitem [{\citenamefont {Swallen}\ \emph {et~al.}(2007)\citenamefont
  {Swallen}, \citenamefont {Kearns}, \citenamefont {Mapes}, \citenamefont
  {Kim}, \citenamefont {McMahon}, \citenamefont {Ediger}, \citenamefont {Wu},
  \citenamefont {Yu},\ and\ \citenamefont {Satija}}]{Ediger2007}%
  \BibitemOpen
  \bibfield  {author} {\bibinfo {author} {\bibfnamefont {S.~F.}\ \bibnamefont
  {Swallen}}, \bibinfo {author} {\bibfnamefont {K.~L.}\ \bibnamefont {Kearns}},
  \bibinfo {author} {\bibfnamefont {M.~K.}\ \bibnamefont {Mapes}}, \bibinfo
  {author} {\bibfnamefont {Y.~S.}\ \bibnamefont {Kim}}, \bibinfo {author}
  {\bibfnamefont {R.~J.}\ \bibnamefont {McMahon}}, \bibinfo {author}
  {\bibfnamefont {M.~D.}\ \bibnamefont {Ediger}}, \bibinfo {author}
  {\bibfnamefont {T.}~\bibnamefont {Wu}}, \bibinfo {author} {\bibfnamefont
  {L.}~\bibnamefont {Yu}}, \ and\ \bibinfo {author} {\bibfnamefont
  {S.}~\bibnamefont {Satija}},\ }\href@noop {} {\bibfield  {journal} {\bibinfo
  {journal} {Science}\ }\textbf {\bibinfo {volume} {315}},\ \bibinfo {pages}
  {353} (\bibinfo {year} {2007})}\BibitemShut {NoStop}%
\bibitem [{\citenamefont {Forrest}\ \emph {et~al.}(1996)\citenamefont
  {Forrest}, \citenamefont {Dalnoki-Veress}, \citenamefont {Stevens},\ and\
  \citenamefont {Dutcher}}]{Forrest1996}%
  \BibitemOpen
  \bibfield  {author} {\bibinfo {author} {\bibfnamefont {J.~A.}\ \bibnamefont
  {Forrest}}, \bibinfo {author} {\bibfnamefont {K.}~\bibnamefont
  {Dalnoki-Veress}}, \bibinfo {author} {\bibfnamefont {J.~R.}\ \bibnamefont
  {Stevens}}, \ and\ \bibinfo {author} {\bibfnamefont {J.~R.}\ \bibnamefont
  {Dutcher}},\ }\href@noop {} {\bibfield  {journal} {\bibinfo  {journal} {Phys.
  Rev. Lett.}\ }\textbf {\bibinfo {volume} {77}},\ \bibinfo {pages} {2002}
  (\bibinfo {year} {1996})}\BibitemShut {NoStop}%
\bibitem [{\citenamefont {Paeng}\ \emph {et~al.}(2011)\citenamefont {Paeng},
  \citenamefont {Swallen},\ and\ \citenamefont {Ediger}}]{Ediger2011}%
  \BibitemOpen
  \bibfield  {author} {\bibinfo {author} {\bibfnamefont {K.}~\bibnamefont
  {Paeng}}, \bibinfo {author} {\bibfnamefont {S.~F.}\ \bibnamefont {Swallen}},
  \ and\ \bibinfo {author} {\bibfnamefont {M.~D.}\ \bibnamefont {Ediger}},\
  }\href@noop {} {\bibfield  {journal} {\bibinfo  {journal} {J. Am. Chem.
  Soc.}\ }\textbf {\bibinfo {volume} {133}},\ \bibinfo {pages} {8444} (\bibinfo
  {year} {2011})}\BibitemShut {NoStop}%
\bibitem [{\citenamefont {Peter}\ \emph {et~al.}(2006)\citenamefont {Peter},
  \citenamefont {Meyer},\ and\ \citenamefont {Baschnagel}}]{Baschnagel2006}%
  \BibitemOpen
  \bibfield  {author} {\bibinfo {author} {\bibfnamefont {S.}~\bibnamefont
  {Peter}}, \bibinfo {author} {\bibfnamefont {H.}~\bibnamefont {Meyer}}, \ and\
  \bibinfo {author} {\bibfnamefont {J.}~\bibnamefont {Baschnagel}},\
  }\href@noop {} {\bibfield  {journal} {\bibinfo  {journal} {J. Pol. Sci., Part
  B: Pol. Phys.}\ }\textbf {\bibinfo {volume} {44}},\ \bibinfo {pages} {2951}
  (\bibinfo {year} {2006})}\BibitemShut {NoStop}%
\bibitem [{\citenamefont {Hanakata}\ \emph {et~al.}(2012)\citenamefont
  {Hanakata}, \citenamefont {Douglas},\ and\ \citenamefont
  {Starr}}]{Douglas2012}%
  \BibitemOpen
  \bibfield  {author} {\bibinfo {author} {\bibfnamefont {P.~Z.}\ \bibnamefont
  {Hanakata}}, \bibinfo {author} {\bibfnamefont {J.~F.}\ \bibnamefont
  {Douglas}}, \ and\ \bibinfo {author} {\bibfnamefont {F.~W.}\ \bibnamefont
  {Starr}},\ }\href@noop {} {\bibfield  {journal} {\bibinfo  {journal} {J.
  Chem. Phys.}\ }\textbf {\bibinfo {volume} {137}},\ \bibinfo {eid} {244901}
  (\bibinfo {year} {2012})}\BibitemShut {NoStop}%
\bibitem [{\citenamefont {Sarabadani}\ \emph {et~al.}(2014)\citenamefont
  {Sarabadani}, \citenamefont {Milchev},\ and\ \citenamefont
  {Vilgis}}]{Vilgis2014}%
  \BibitemOpen
  \bibfield  {author} {\bibinfo {author} {\bibfnamefont {J.}~\bibnamefont
  {Sarabadani}}, \bibinfo {author} {\bibfnamefont {A.}~\bibnamefont {Milchev}},
  \ and\ \bibinfo {author} {\bibfnamefont {T.~A.}\ \bibnamefont {Vilgis}},\
  }\href@noop {} {\bibfield  {journal} {\bibinfo  {journal} {J. Chem. Phys.}\
  }\textbf {\bibinfo {volume} {141}},\ \bibinfo {eid} {044907} (\bibinfo {year}
  {2014})}\BibitemShut {NoStop}%
\bibitem [{\citenamefont {White}\ \emph {et~al.}(2015)\citenamefont {White},
  \citenamefont {Price},\ and\ \citenamefont {Lipson}}]{Lipson2015}%
  \BibitemOpen
  \bibfield  {author} {\bibinfo {author} {\bibfnamefont {R.~P.}\ \bibnamefont
  {White}}, \bibinfo {author} {\bibfnamefont {C.~C.}\ \bibnamefont {Price}}, \
  and\ \bibinfo {author} {\bibfnamefont {J.~E.~G.}\ \bibnamefont {Lipson}},\
  }\href@noop {} {\bibfield  {journal} {\bibinfo  {journal} {Macromolecules}\
  }\textbf {\bibinfo {volume} {48}},\ \bibinfo {pages} {4132} (\bibinfo {year}
  {2015})}\BibitemShut {NoStop}%
\bibitem [{\citenamefont {Inoue}\ \emph {et~al.}(2011)\citenamefont {Inoue},
  \citenamefont {Kawashima}, \citenamefont {Matsui}, \citenamefont {Kanaya},
  \citenamefont {Nishida}, \citenamefont {Matsuba},\ and\ \citenamefont
  {Hino}}]{Kanaya2011}%
  \BibitemOpen
  \bibfield  {author} {\bibinfo {author} {\bibfnamefont {R.}~\bibnamefont
  {Inoue}}, \bibinfo {author} {\bibfnamefont {K.}~\bibnamefont {Kawashima}},
  \bibinfo {author} {\bibfnamefont {K.}~\bibnamefont {Matsui}}, \bibinfo
  {author} {\bibfnamefont {T.}~\bibnamefont {Kanaya}}, \bibinfo {author}
  {\bibfnamefont {K.}~\bibnamefont {Nishida}}, \bibinfo {author} {\bibfnamefont
  {G.}~\bibnamefont {Matsuba}}, \ and\ \bibinfo {author} {\bibfnamefont
  {M.}~\bibnamefont {Hino}},\ }\href@noop {} {\bibfield  {journal} {\bibinfo
  {journal} {Phys. Rev. E}\ }\textbf {\bibinfo {volume} {83}},\ \bibinfo
  {pages} {021801} (\bibinfo {year} {2011})}\BibitemShut {NoStop}%
\bibitem [{\citenamefont {SchÃ¶nherr}\ and\ \citenamefont
  {Frank}(2003)}]{Frank2003}%
  \BibitemOpen
  \bibfield  {author} {\bibinfo {author} {\bibfnamefont {H.}~\bibnamefont
  {SchÃ¶nherr}}\ and\ \bibinfo {author} {\bibfnamefont {C.~W.}\ \bibnamefont
  {Frank}},\ }\href@noop {} {\bibfield  {journal} {\bibinfo  {journal}
  {Macromolecules}\ }\textbf {\bibinfo {volume} {36}},\ \bibinfo {pages} {1199}
  (\bibinfo {year} {2003})}\BibitemShut {NoStop}%
\bibitem [{\citenamefont {Spiece}\ \emph {et~al.}(2015)\citenamefont {Spiece},
  \citenamefont {Martinez-Tong}, \citenamefont {Sferrazza}, \citenamefont
  {Nogales},\ and\ \citenamefont {Napolitano}}]{Napolitano2015}%
  \BibitemOpen
  \bibfield  {author} {\bibinfo {author} {\bibfnamefont {J.}~\bibnamefont
  {Spiece}}, \bibinfo {author} {\bibfnamefont {D.~E.}\ \bibnamefont
  {Martinez-Tong}}, \bibinfo {author} {\bibfnamefont {M.}~\bibnamefont
  {Sferrazza}}, \bibinfo {author} {\bibfnamefont {A.}~\bibnamefont {Nogales}},
  \ and\ \bibinfo {author} {\bibfnamefont {S.}~\bibnamefont {Napolitano}},\
  }\href {\doibase 10.1039/C5SM01229E} {\bibfield  {journal} {\bibinfo
  {journal} {Soft Matt.}\ }\textbf {\bibinfo {volume} {11}},\ \bibinfo {pages}
  {6179} (\bibinfo {year} {2015})}\BibitemShut {NoStop}%
\bibitem [{\citenamefont {Alcoutlabi}\ and\ \citenamefont
  {McKenna}(2005)}]{McKenna2005}%
  \BibitemOpen
  \bibfield  {author} {\bibinfo {author} {\bibfnamefont {M.}~\bibnamefont
  {Alcoutlabi}}\ and\ \bibinfo {author} {\bibfnamefont {G.~B.}\ \bibnamefont
  {McKenna}},\ }\href {http://stacks.iop.org/0953-8984/17/i=15/a=R01}
  {\bibfield  {journal} {\bibinfo  {journal} {J. Phys.: Cond. Matt.}\ }\textbf
  {\bibinfo {volume} {17}},\ \bibinfo {pages} {R461} (\bibinfo {year}
  {2005})}\BibitemShut {NoStop}%
\bibitem [{\citenamefont {Napolitano}\ \emph {et~al.}(2012)\citenamefont
  {Napolitano}, \citenamefont {Rotella},\ and\ \citenamefont
  {Wubbenhorst}}]{Napolitano2012}%
  \BibitemOpen
  \bibfield  {author} {\bibinfo {author} {\bibfnamefont {S.}~\bibnamefont
  {Napolitano}}, \bibinfo {author} {\bibfnamefont {C.}~\bibnamefont {Rotella}},
  \ and\ \bibinfo {author} {\bibfnamefont {M.}~\bibnamefont {Wubbenhorst}},\
  }\href@noop {} {\bibfield  {journal} {\bibinfo  {journal} {ACS Macro Lett.}\
  }\textbf {\bibinfo {volume} {1}},\ \bibinfo {pages} {1189} (\bibinfo {year}
  {2012})}\BibitemShut {NoStop}%
\bibitem [{\citenamefont {Cangialosi}\ \emph {et~al.}(2016)\citenamefont
  {Cangialosi}, \citenamefont {Alegria},\ and\ \citenamefont
  {Colmenero}}]{Cangialosi2015}%
  \BibitemOpen
  \bibfield  {author} {\bibinfo {author} {\bibfnamefont {D.}~\bibnamefont
  {Cangialosi}}, \bibinfo {author} {\bibfnamefont {A.}~\bibnamefont {Alegria}},
  \ and\ \bibinfo {author} {\bibfnamefont {J.}~\bibnamefont {Colmenero}},\
  }\href@noop {} {\bibfield  {journal} {\bibinfo  {journal} {Prog. Pol. Sci.}\
  }\textbf {\bibinfo {volume} {54â€“55}},\ \bibinfo {pages} {128 } (\bibinfo
  {year} {2016})}\BibitemShut {NoStop}%
\bibitem [{\citenamefont {Kremer}\ \emph {et~al.}(2015)\citenamefont {Kremer},
  \citenamefont {Tress},\ and\ \citenamefont {Mapesa}}]{Kremer2015}%
  \BibitemOpen
  \bibfield  {author} {\bibinfo {author} {\bibfnamefont {F.}~\bibnamefont
  {Kremer}}, \bibinfo {author} {\bibfnamefont {M.}~\bibnamefont {Tress}}, \
  and\ \bibinfo {author} {\bibfnamefont {E.~U.}\ \bibnamefont {Mapesa}},\
  }\href {\doibase http://dx.doi.org/10.1016/j.jnoncrysol.2014.08.016}
  {\bibfield  {journal} {\bibinfo  {journal} {J. Non-Cryst. Sol.}\ }\textbf
  {\bibinfo {volume} {407}},\ \bibinfo {pages} {277 } (\bibinfo {year}
  {2015})}\BibitemShut {NoStop}%
\bibitem [{\citenamefont {Priestley}\ \emph {et~al.}(2015)\citenamefont
  {Priestley}, \citenamefont {Cangialosi},\ and\ \citenamefont
  {Napolitano}}]{Priestley2015}%
  \BibitemOpen
  \bibfield  {author} {\bibinfo {author} {\bibfnamefont {R.~D.}\ \bibnamefont
  {Priestley}}, \bibinfo {author} {\bibfnamefont {D.}~\bibnamefont
  {Cangialosi}}, \ and\ \bibinfo {author} {\bibfnamefont {S.}~\bibnamefont
  {Napolitano}},\ }\href {\doibase
  http://dx.doi.org/10.1016/j.jnoncrysol.2014.09.048} {\bibfield  {journal}
  {\bibinfo  {journal} {J. Non-Cryst. Sol.}\ }\textbf {\bibinfo {volume}
  {407}},\ \bibinfo {pages} {288 } (\bibinfo {year} {2015})}\BibitemShut
  {NoStop}%
\bibitem [{\citenamefont {Boucher}\ \emph
  {et~al.}(2012{\natexlab{a}})\citenamefont {Boucher}, \citenamefont
  {Cangialosi}, \citenamefont {Alegr\'{i}a},\ and\ \citenamefont
  {Colmenero}}]{Boucher2012}%
  \BibitemOpen
  \bibfield  {author} {\bibinfo {author} {\bibfnamefont {V.~M.}\ \bibnamefont
  {Boucher}}, \bibinfo {author} {\bibfnamefont {D.}~\bibnamefont {Cangialosi}},
  \bibinfo {author} {\bibfnamefont {A.}~\bibnamefont {Alegr\'{i}a}}, \ and\
  \bibinfo {author} {\bibfnamefont {J.}~\bibnamefont {Colmenero}},\ }\href@noop
  {} {\bibfield  {journal} {\bibinfo  {journal} {Macromolecules}\ }\textbf
  {\bibinfo {volume} {45}},\ \bibinfo {pages} {5296} (\bibinfo {year}
  {2012}{\natexlab{a}})}\BibitemShut {NoStop}%
\bibitem [{\citenamefont {Cangialosi}\ \emph {et~al.}(2013)\citenamefont
  {Cangialosi}, \citenamefont {Boucher}, \citenamefont {Alegr\'{\i}a},\ and\
  \citenamefont {Colmenero}}]{Cangialosi2013}%
  \BibitemOpen
  \bibfield  {author} {\bibinfo {author} {\bibfnamefont {D.}~\bibnamefont
  {Cangialosi}}, \bibinfo {author} {\bibfnamefont {V.~M.}\ \bibnamefont
  {Boucher}}, \bibinfo {author} {\bibfnamefont {A.}~\bibnamefont
  {Alegr\'{\i}a}}, \ and\ \bibinfo {author} {\bibfnamefont {J.}~\bibnamefont
  {Colmenero}},\ }\href@noop {} {\bibfield  {journal} {\bibinfo  {journal}
  {Phys. Rev. Lett.}\ }\textbf {\bibinfo {volume} {111}},\ \bibinfo {pages}
  {095701} (\bibinfo {year} {2013})}\BibitemShut {NoStop}%
\bibitem [{\citenamefont {Pye}\ and\ \citenamefont {Roth}(2011)}]{Roth2011}%
  \BibitemOpen
  \bibfield  {author} {\bibinfo {author} {\bibfnamefont {J.~E.}\ \bibnamefont
  {Pye}}\ and\ \bibinfo {author} {\bibfnamefont {C.~B.}\ \bibnamefont {Roth}},\
  }\href@noop {} {\bibfield  {journal} {\bibinfo  {journal} {Phys. Rev. Lett.}\
  }\textbf {\bibinfo {volume} {107}},\ \bibinfo {pages} {235701} (\bibinfo
  {year} {2011})}\BibitemShut {NoStop}%
\bibitem [{\citenamefont {Wisitsorasak}\ and\ \citenamefont
  {Wolynes}(2014)}]{Wolynes2014}%
  \BibitemOpen
  \bibfield  {author} {\bibinfo {author} {\bibfnamefont {A.}~\bibnamefont
  {Wisitsorasak}}\ and\ \bibinfo {author} {\bibfnamefont {P.~G.}\ \bibnamefont
  {Wolynes}},\ }\href@noop {} {\bibfield  {journal} {\bibinfo  {journal} {J.
  Phys. Chem. B}\ }\textbf {\bibinfo {volume} {118}},\ \bibinfo {pages} {7835}
  (\bibinfo {year} {2014})}\BibitemShut {NoStop}%
\bibitem [{\citenamefont {Reiter}\ \emph {et~al.}(2005)\citenamefont {Reiter},
  \citenamefont {Hamieh}, \citenamefont {Damman}, \citenamefont {Sclavons},
  \citenamefont {Gabriele}, \citenamefont {Vilmin},\ and\ \citenamefont
  {Raphael}}]{Reiter2005}%
  \BibitemOpen
  \bibfield  {author} {\bibinfo {author} {\bibfnamefont {G.}~\bibnamefont
  {Reiter}}, \bibinfo {author} {\bibfnamefont {M.}~\bibnamefont {Hamieh}},
  \bibinfo {author} {\bibfnamefont {P.}~\bibnamefont {Damman}}, \bibinfo
  {author} {\bibfnamefont {S.}~\bibnamefont {Sclavons}}, \bibinfo {author}
  {\bibfnamefont {S.}~\bibnamefont {Gabriele}}, \bibinfo {author}
  {\bibfnamefont {T.}~\bibnamefont {Vilmin}}, \ and\ \bibinfo {author}
  {\bibfnamefont {E.}~\bibnamefont {Raphael}},\ }\href@noop {} {\bibfield
  {journal} {\bibinfo  {journal} {Nat. Mater.}\ }\textbf {\bibinfo {volume}
  {4}},\ \bibinfo {pages} {754} (\bibinfo {year} {2005})}\BibitemShut {NoStop}%
\bibitem [{\citenamefont {Serghei}\ and\ \citenamefont
  {Kremer}(2008)}]{Kremer2008}%
  \BibitemOpen
  \bibfield  {author} {\bibinfo {author} {\bibfnamefont {A.}~\bibnamefont
  {Serghei}}\ and\ \bibinfo {author} {\bibfnamefont {F.}~\bibnamefont
  {Kremer}},\ }\href@noop {} {\bibfield  {journal} {\bibinfo  {journal}
  {Macromol. Chem. Phys.}\ }\textbf {\bibinfo {volume} {209}},\ \bibinfo
  {pages} {810} (\bibinfo {year} {2008})}\BibitemShut {NoStop}%
\bibitem [{\citenamefont {Whitlow}\ and\ \citenamefont
  {Wool}(1991)}]{Wool1991}%
  \BibitemOpen
  \bibfield  {author} {\bibinfo {author} {\bibfnamefont {S.~J.}\ \bibnamefont
  {Whitlow}}\ and\ \bibinfo {author} {\bibfnamefont {R.~P.}\ \bibnamefont
  {Wool}},\ }\href@noop {} {\bibfield  {journal} {\bibinfo  {journal}
  {Macromolecules}\ }\textbf {\bibinfo {volume} {24}},\ \bibinfo {pages} {5926}
  (\bibinfo {year} {1991})}\BibitemShut {NoStop}%
\bibitem [{\citenamefont {Boucher}\ \emph
  {et~al.}(2012{\natexlab{b}})\citenamefont {Boucher}, \citenamefont
  {Cangialosi}, \citenamefont {Yin}, \citenamefont {Schonhals}, \citenamefont
  {Alegria},\ and\ \citenamefont {Colmenero}}]{Cangialosi2012}%
  \BibitemOpen
  \bibfield  {author} {\bibinfo {author} {\bibfnamefont {V.~M.}\ \bibnamefont
  {Boucher}}, \bibinfo {author} {\bibfnamefont {D.}~\bibnamefont {Cangialosi}},
  \bibinfo {author} {\bibfnamefont {H.}~\bibnamefont {Yin}}, \bibinfo {author}
  {\bibfnamefont {A.}~\bibnamefont {Schonhals}}, \bibinfo {author}
  {\bibfnamefont {A.}~\bibnamefont {Alegria}}, \ and\ \bibinfo {author}
  {\bibfnamefont {J.}~\bibnamefont {Colmenero}},\ }\href@noop {} {\bibfield
  {journal} {\bibinfo  {journal} {Soft Matt.}\ }\textbf {\bibinfo {volume}
  {8}},\ \bibinfo {pages} {5119} (\bibinfo {year}
  {2012}{\natexlab{b}})}\BibitemShut {NoStop}%
\bibitem [{\citenamefont {Moynihan}\ \emph {et~al.}(1976)\citenamefont
  {Moynihan}, \citenamefont {Macedo}, \citenamefont {Montrose}, \citenamefont
  {Gupta}, \citenamefont {De~Bolt}, \citenamefont {Dill}, \citenamefont {Dom},
  \citenamefont {Drake}, \citenamefont {Eastel}, \citenamefont {Elterman},
  \citenamefont {Moeller}, \citenamefont {Sasabe},\ and\ \citenamefont
  {Wilder}}]{Moynihan1976}%
  \BibitemOpen
  \bibfield  {author} {\bibinfo {author} {\bibfnamefont {C.~T.}\ \bibnamefont
  {Moynihan}}, \bibinfo {author} {\bibfnamefont {P.~B.}\ \bibnamefont
  {Macedo}}, \bibinfo {author} {\bibfnamefont {C.~J.}\ \bibnamefont
  {Montrose}}, \bibinfo {author} {\bibfnamefont {P.~K.}\ \bibnamefont {Gupta}},
  \bibinfo {author} {\bibfnamefont {M.~A.}\ \bibnamefont {De~Bolt}}, \bibinfo
  {author} {\bibfnamefont {J.~F.}\ \bibnamefont {Dill}}, \bibinfo {author}
  {\bibfnamefont {B.~E.}\ \bibnamefont {Dom}}, \bibinfo {author} {\bibfnamefont
  {P.~W.}\ \bibnamefont {Drake}}, \bibinfo {author} {\bibfnamefont {A.~J.}\
  \bibnamefont {Eastel}}, \bibinfo {author} {\bibfnamefont {P.~B.}\
  \bibnamefont {Elterman}}, \bibinfo {author} {\bibfnamefont {R.~P.}\
  \bibnamefont {Moeller}}, \bibinfo {author} {\bibfnamefont {H.}~\bibnamefont
  {Sasabe}}, \ and\ \bibinfo {author} {\bibfnamefont {J.~A.}\ \bibnamefont
  {Wilder}},\ }\href@noop {} {\bibfield  {journal} {\bibinfo  {journal} {Ann.
  NY Acad. Sci.}\ }\textbf {\bibinfo {volume} {279}},\ \bibinfo {pages} {15}
  (\bibinfo {year} {1976})}\BibitemShut {NoStop}%
\bibitem [{\citenamefont {Miller}(1970)}]{Miller1970}%
  \BibitemOpen
  \bibfield  {author} {\bibinfo {author} {\bibfnamefont {A.~A.}\ \bibnamefont
  {Miller}},\ }\href@noop {} {\bibfield  {journal} {\bibinfo  {journal}
  {Macromolecules}\ }\textbf {\bibinfo {volume} {3}},\ \bibinfo {pages} {674}
  (\bibinfo {year} {1970})}\BibitemShut {NoStop}%
\bibitem [{\citenamefont {Hodge}\ and\ \citenamefont
  {O'Reilly}(1999)}]{Hodge1999}%
  \BibitemOpen
  \bibfield  {author} {\bibinfo {author} {\bibfnamefont {I.~M.}\ \bibnamefont
  {Hodge}}\ and\ \bibinfo {author} {\bibfnamefont {J.~M.}\ \bibnamefont
  {O'Reilly}},\ }\href@noop {} {\bibfield  {journal} {\bibinfo  {journal} {J.
  Phys. Chem. B}\ }\textbf {\bibinfo {volume} {103}},\ \bibinfo {pages} {4171}
  (\bibinfo {year} {1999})}\BibitemShut {NoStop}%
\bibitem [{\citenamefont {Cangialosi}\ \emph {et~al.}(2005)\citenamefont
  {Cangialosi}, \citenamefont {Alegria},\ and\ \citenamefont
  {Colmenero}}]{Cangialosi2005}%
  \BibitemOpen
  \bibfield  {author} {\bibinfo {author} {\bibfnamefont {D.}~\bibnamefont
  {Cangialosi}}, \bibinfo {author} {\bibfnamefont {A.}~\bibnamefont {Alegria}},
  \ and\ \bibinfo {author} {\bibfnamefont {J.}~\bibnamefont {Colmenero}},\
  }\href@noop {} {\bibfield  {journal} {\bibinfo  {journal} {Europhys. Lett.}\
  }\textbf {\bibinfo {volume} {70}},\ \bibinfo {pages} {614} (\bibinfo {year}
  {2005})}\BibitemShut {NoStop}%
\bibitem [{\citenamefont {Pyda}(2003)}]{ATHAS}%
  \BibitemOpen
  \bibfield  {author} {\bibinfo {author} {\bibfnamefont {M.}~\bibnamefont
  {Pyda}},\ }\href@noop {} {\bibfield  {journal} {\bibinfo  {journal} {ATHAS
  Data Bank}\ ,\ \bibinfo {pages}
  {http://web.utk.edu/~athas/databank/intro.html}} (\bibinfo {year}
  {2003})}\BibitemShut {NoStop}%
\bibitem [{Note1()}]{Note1}%
  \BibitemOpen
  \bibinfo {note} {Data for crystalline PS, though not experimentally available
  due to the inability of this polymer to crystallize, are either obtained
  theoretically or relying on the ability of isotactic PS to crystallize. In
  this case, as shown by O'Reilly and coworkers \cite {OReilly1965}, the lack
  of effect of tacticity on equilibrium thermodynamic properties is
  exploited.}\BibitemShut {Stop}%
\bibitem [{\citenamefont {Stillinger}\ and\ \citenamefont
  {Debenedetti}(2013)}]{Debenedetti2013}%
  \BibitemOpen
  \bibfield  {author} {\bibinfo {author} {\bibfnamefont {F.~H.}\ \bibnamefont
  {Stillinger}}\ and\ \bibinfo {author} {\bibfnamefont {P.~G.}\ \bibnamefont
  {Debenedetti}},\ }\href@noop {} {\bibfield  {journal} {\bibinfo  {journal}
  {Ann. Rev. Cond. Matt. Phys.}\ }\textbf {\bibinfo {volume} {4}},\ \bibinfo
  {pages} {263} (\bibinfo {year} {2013})}\BibitemShut {NoStop}%
\bibitem [{\citenamefont {Kirkpatrick}\ \emph {et~al.}(1989)\citenamefont
  {Kirkpatrick}, \citenamefont {Thirumalai},\ and\ \citenamefont
  {Wolynes}}]{Wolynes1989}%
  \BibitemOpen
  \bibfield  {author} {\bibinfo {author} {\bibfnamefont {T.~R.}\ \bibnamefont
  {Kirkpatrick}}, \bibinfo {author} {\bibfnamefont {D.}~\bibnamefont
  {Thirumalai}}, \ and\ \bibinfo {author} {\bibfnamefont {P.~G.}\ \bibnamefont
  {Wolynes}},\ }\href@noop {} {\bibfield  {journal} {\bibinfo  {journal} {Phys.
  Rev. A}\ }\textbf {\bibinfo {volume} {40}},\ \bibinfo {pages} {1045}
  (\bibinfo {year} {1989})}\BibitemShut {NoStop}%
\bibitem [{\citenamefont {M\'ezard}\ and\ \citenamefont
  {Parisi}(1999)}]{Parisi1999}%
  \BibitemOpen
  \bibfield  {author} {\bibinfo {author} {\bibfnamefont {M.}~\bibnamefont
  {M\'ezard}}\ and\ \bibinfo {author} {\bibfnamefont {G.}~\bibnamefont
  {Parisi}},\ }\href@noop {} {\bibfield  {journal} {\bibinfo  {journal} {Phys.
  Rev. Lett.}\ }\textbf {\bibinfo {volume} {82}},\ \bibinfo {pages} {747}
  (\bibinfo {year} {1999})}\BibitemShut {NoStop}%
\bibitem [{\citenamefont {Cammarota}\ and\ \citenamefont
  {Biroli}(2012)}]{Cammarota2012}%
  \BibitemOpen
  \bibfield  {author} {\bibinfo {author} {\bibfnamefont {C.}~\bibnamefont
  {Cammarota}}\ and\ \bibinfo {author} {\bibfnamefont {G.}~\bibnamefont
  {Biroli}},\ }\href {\doibase 10.1073/pnas.1111582109} {\bibfield  {journal}
  {\bibinfo  {journal} {Proc. Natl. Acad. Sci. USA}\ }\textbf {\bibinfo
  {volume} {109}},\ \bibinfo {pages} {8850} (\bibinfo {year}
  {2012})}\BibitemShut {NoStop}%
\bibitem [{\citenamefont {Ozawa}\ \emph {et~al.}(2015)\citenamefont {Ozawa},
  \citenamefont {Kob}, \citenamefont {Ikeda},\ and\ \citenamefont
  {Miyazaki}}]{Kob2015}%
  \BibitemOpen
  \bibfield  {author} {\bibinfo {author} {\bibfnamefont {M.}~\bibnamefont
  {Ozawa}}, \bibinfo {author} {\bibfnamefont {W.}~\bibnamefont {Kob}}, \bibinfo
  {author} {\bibfnamefont {A.}~\bibnamefont {Ikeda}}, \ and\ \bibinfo {author}
  {\bibfnamefont {K.}~\bibnamefont {Miyazaki}},\ }\href {\doibase
  10.1073/pnas.1500730112} {\bibfield  {journal} {\bibinfo  {journal} {Proc.
  Natl. Acad. Sci. USA}\ }\textbf {\bibinfo {volume} {112}},\ \bibinfo {pages}
  {6914} (\bibinfo {year} {2015})}\BibitemShut {NoStop}%
\bibitem [{\citenamefont {Hecksher}\ \emph {et~al.}(2008)\citenamefont
  {Hecksher}, \citenamefont {Nielsen}, \citenamefont {Olsen},\ and\
  \citenamefont {Dyre}}]{Dyre2008}%
  \BibitemOpen
  \bibfield  {author} {\bibinfo {author} {\bibfnamefont {T.}~\bibnamefont
  {Hecksher}}, \bibinfo {author} {\bibfnamefont {A.~I.}\ \bibnamefont
  {Nielsen}}, \bibinfo {author} {\bibfnamefont {N.~B.}\ \bibnamefont {Olsen}},
  \ and\ \bibinfo {author} {\bibfnamefont {J.~C.}\ \bibnamefont {Dyre}},\
  }\href@noop {} {\bibfield  {journal} {\bibinfo  {journal} {Nat. Phys.}\
  }\textbf {\bibinfo {volume} {4}},\ \bibinfo {pages} {737} (\bibinfo {year}
  {2008})}\BibitemShut {NoStop}%
\bibitem [{\citenamefont {Pogna}\ \emph {et~al.}(2015)\citenamefont {Pogna},
  \citenamefont {RodrÃ­guez-Tinoco}, \citenamefont {Cerullo}, \citenamefont
  {Ferrante}, \citenamefont {RodrÃ­guez-Viejo},\ and\ \citenamefont
  {Scopigno}}]{Scopigno2015}%
  \BibitemOpen
  \bibfield  {author} {\bibinfo {author} {\bibfnamefont {E.~A.~A.}\
  \bibnamefont {Pogna}}, \bibinfo {author} {\bibfnamefont {C.}~\bibnamefont
  {RodrÃ­guez-Tinoco}}, \bibinfo {author} {\bibfnamefont {G.}~\bibnamefont
  {Cerullo}}, \bibinfo {author} {\bibfnamefont {C.}~\bibnamefont {Ferrante}},
  \bibinfo {author} {\bibfnamefont {J.}~\bibnamefont {RodrÃ­guez-Viejo}}, \
  and\ \bibinfo {author} {\bibfnamefont {T.}~\bibnamefont {Scopigno}},\
  }\href@noop {} {\bibfield  {journal} {\bibinfo  {journal} {Proc. Natl. Acad.
  Sci. USA}\ }\textbf {\bibinfo {volume} {112}},\ \bibinfo {pages} {2331}
  (\bibinfo {year} {2015})}\BibitemShut {NoStop}%
\bibitem [{\citenamefont {Chakrabarty}\ \emph {et~al.}(2015)\citenamefont
  {Chakrabarty}, \citenamefont {Karmakar},\ and\ \citenamefont
  {Dasgupta}}]{Chakrabarty2015}%
  \BibitemOpen
  \bibfield  {author} {\bibinfo {author} {\bibfnamefont {S.}~\bibnamefont
  {Chakrabarty}}, \bibinfo {author} {\bibfnamefont {S.}~\bibnamefont
  {Karmakar}}, \ and\ \bibinfo {author} {\bibfnamefont {C.}~\bibnamefont
  {Dasgupta}},\ }\href@noop {} {\bibfield  {journal} {\bibinfo  {journal}
  {Proc. Natl. Acad. Sci. USA}\ }\textbf {\bibinfo {volume} {112}},\ \bibinfo
  {pages} {E4819} (\bibinfo {year} {2015})}\BibitemShut {NoStop}%
\bibitem [{\citenamefont {Karasz}\ \emph {et~al.}(1965)\citenamefont {Karasz},
  \citenamefont {Bair},\ and\ \citenamefont {O'Reilly}}]{OReilly1965}%
  \BibitemOpen
  \bibfield  {author} {\bibinfo {author} {\bibfnamefont {F.~E.}\ \bibnamefont
  {Karasz}}, \bibinfo {author} {\bibfnamefont {H.~E.}\ \bibnamefont {Bair}}, \
  and\ \bibinfo {author} {\bibfnamefont {J.~M.}\ \bibnamefont {O'Reilly}},\
  }\href@noop {} {\bibfield  {journal} {\bibinfo  {journal} {J. Phys. Chem.}\
  }\textbf {\bibinfo {volume} {69}},\ \bibinfo {pages} {2657} (\bibinfo {year}
  {1965})}\BibitemShut {NoStop}%
\end{thebibliography}
\end{document}


\preprint{APS/123-QED}

\title{Supplementary material for: Reaching the ideal glass transition by aging polymer films}

\author{Virginie M. Boucher$^1$}

\author{Daniele Cangialosi$^1$}%
 \email{Second.Author@institution.edu}
 \author{Angel Alegr\'{\i}a$^{1,2}$}
 \author{Juan Colmenero$^{1,2,3}$}

\affiliation{%
 $^1$Centro de F\'{\i}sica de Materiales, Paseo Manuel de Lardizabal 5, 20018 San Sebasti\'{a}n, Spain}%

 \homepage{http://www.Second.institution.edu/~Charlie.Author}
\affiliation{
 $^2$Departamento de F\'{\i}sica de Materiales (UPV/EHU), Apartado 1072, 20080 San Sebasti\'{a}n, Spain\\
}%
\affiliation{
 $^3$Donostia International Physics Center, Paseo Manuel de Lardizabal 4, 20018 San Sebasti\'{a}n, Spain
}%

\date{\today}

\begin{abstract}

\end{abstract}

\pacs{Valid PACS appear here}
\maketitle

\begin{figure} [htb!]
\centering
\includegraphics[width=.45\textwidth, angle=0]{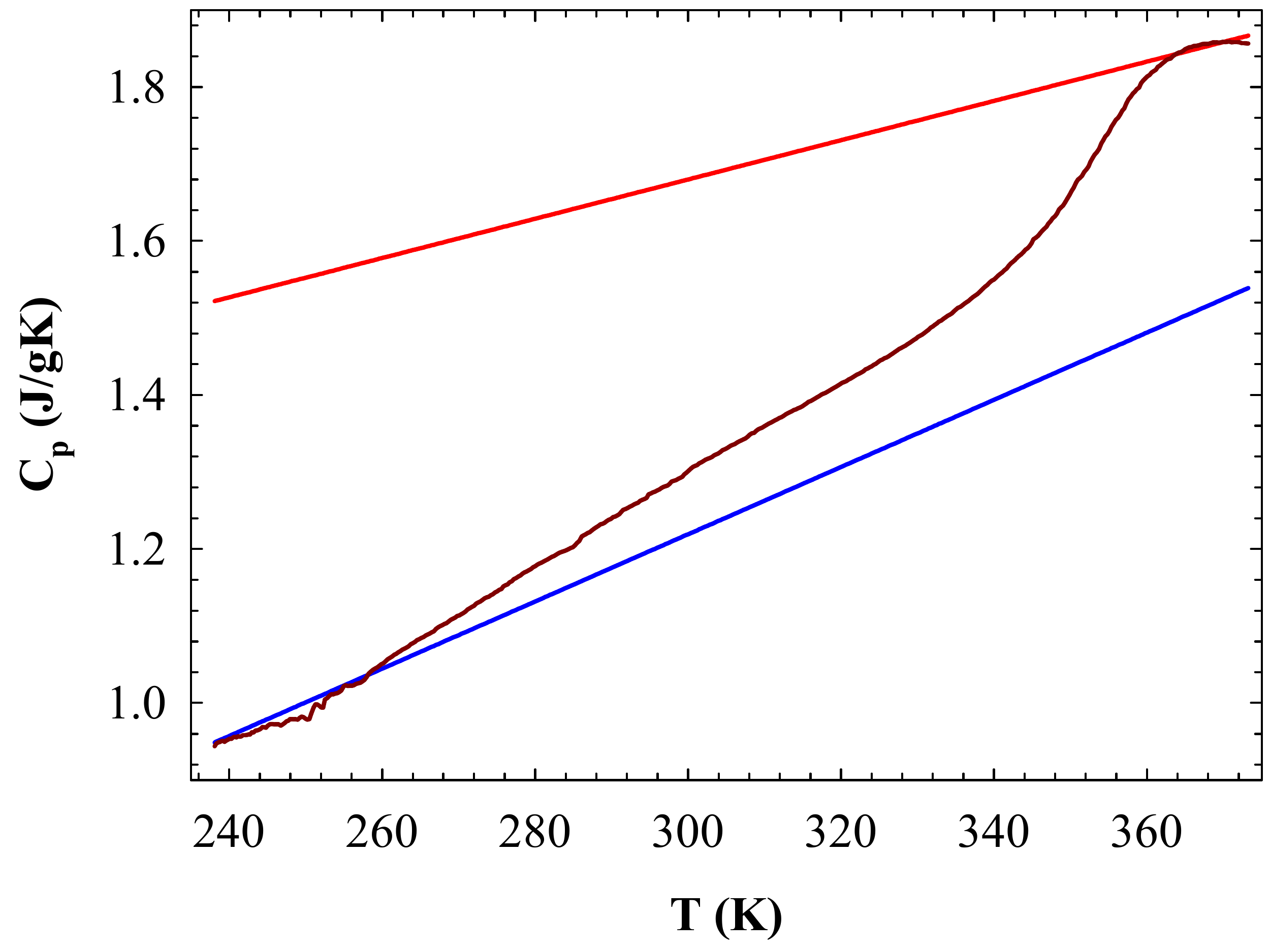}
\caption{Heat capacity as a function of temperature for 30 nm PS films annealed at 248 K at the indicated aging time. The reference curve corresponds to a sample heated up just after cooling. The inset shows the resulting recovered enthalpies (right axis) and $T_f$(s) (left axis) as a function of the aging time.}
\label{Fig2SI}
\end{figure}

{\bf Isothermal enthalpy recovery} As a showcase, DSC heating scans recorded after different aging times at 248 K are shown in Figure \ref{Fig2SI}. A pronounced broad overshoot, increasing in intensity with the aging time, is evident. The inset of Figure \ref{Fig2SI} shows the aging time dependence of the recovered enthalpy and the corresponding values of $T_f$. At the longest aging times the amount of recovered enthalpy is about 23 J/g. In such time scale, that is, several days, this amount is considerably larger than normally observed in conventional glasses. $T_f$ reaches a plateau value of 295 K, after the longest aging time.
\\

\begin{figure} [htb!]
\centering
\includegraphics[width=.45\textwidth, angle=0]{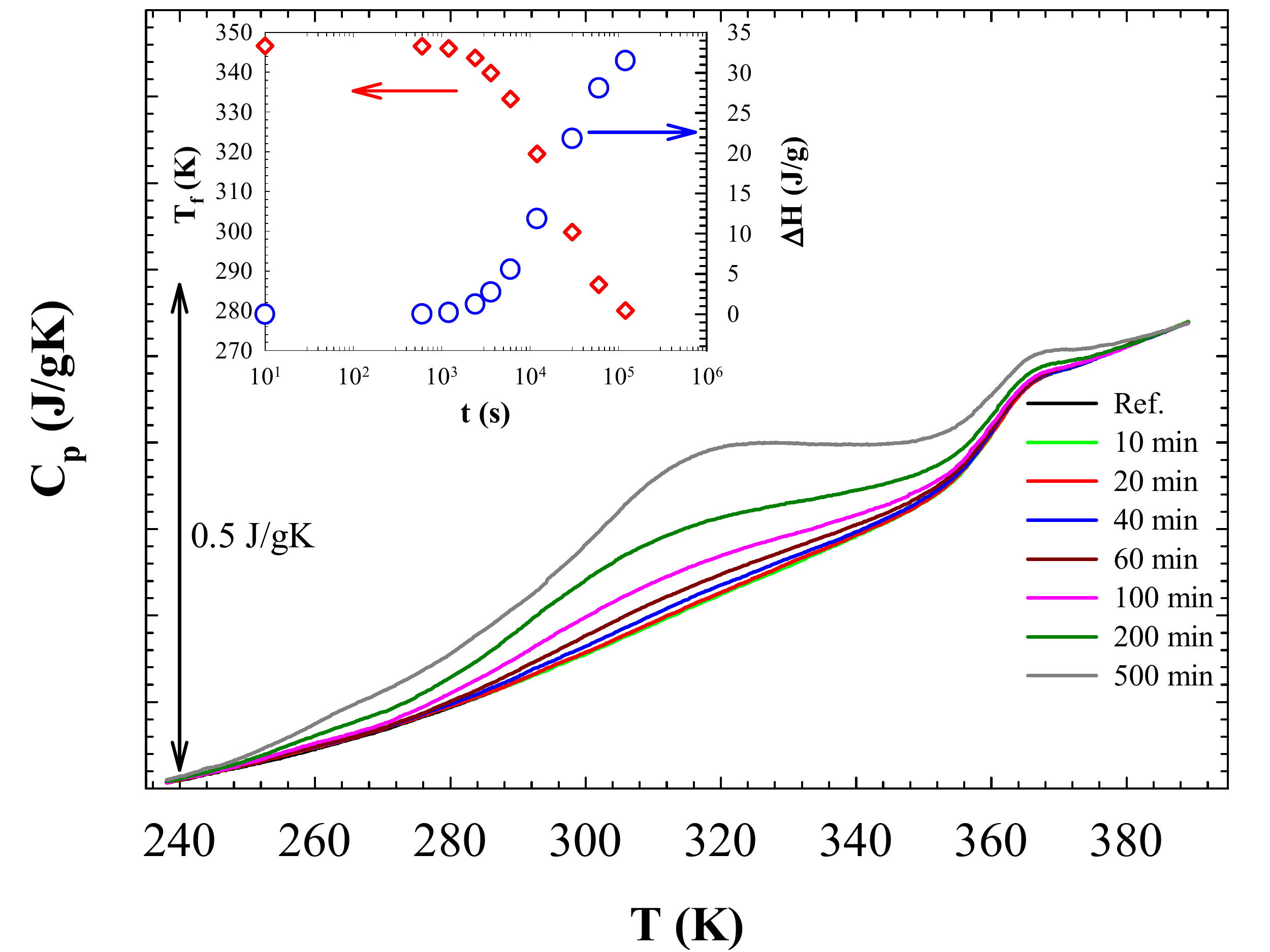}
\caption{Heat capacity as a function of temperature of 30 nm thick PS films cooled down at 20 K/min (brown line), melt (red line) and glass (blue line) bulk PS.}
\label{Fig1SI}
\end{figure}

{\bf Specific heat: 30 nm films vs. bulk} An important aspect of the present study regards the equivalence of the equilibrium thermodynamics of 30 nm PS films with the bulk polymer. Hence, in order to define the thermodynamic diagram when cooling 30 nm PS films at 20 K/min, the specific heat was determined. This is shown in Figure \ref{Fig1SI} (brown line) after correction of the baseline. The experimental specific heat is compared to the glass and melt specific heats reported in the ATHAS databank \cite{ATHAS}, also shown in Figure \ref{Fig1SI} (blue and red lines respectively). As can be seen, once the specific heat of the present study of the glass and the melt are taken at sufficiently low and high temperatures respectively (below $\sim$ 260 K and above $\sim$ 355 K), the 30 nm thick sample exhibits melt and glass specific heats equal to those of bulk PS. The former result agrees with previous studies showing that the equilibrium thermodynamics is essentially bulk-like for freestanding films as thin as 30 nm \cite{Baschnagel2006,Douglas2012,Vilgis2014,Lipson2015,Kanaya2011}.

{\bf Two steps recovery of equilibrium} The presence of two molecular mechanisms driving the 30 nm thick PS toward equilibrium and its analogy with bulk glass formers constitutes a crucial ingredient to achieve the ideal glass transition. In bulk glass formers, this has been recently found studying the isothermal enthalpy recovery in several polymers \cite{Cangialosi2013}. Achieving information on the presence of multiple mechanisms, and their corresponding thermodynamic states in metastable equilibrium, generally requires experimentally demanding observation time scales. However, there exist numerous experimental studies on the enthalpy recovery of different glasses, showing a plateau with partial recovery \cite{Bauwens-Crowet1986,Cowie1997,Duenas1997,Hutchinson2002,Andreozzi2003,Cowie2005,Boucher2011,KohPydaSimon2013,Tanaka2016}. In all these cases, the amount of recoverable enthalpy, determined from the knowledge of the melt and glass specific heats $-$ providing the specific heat jump $-$ and the distance from the glass transition ($T_g$ $-$ $T_a$), is $always$ considerably larger than the enthalpy actually recovered at a plateau. Importantly, this is the case even in one study where the total amount of recoverable enthalpy is erroneously evaluated \cite{Hutchinson2002} due to an underestimation of the specific heat jump: $\Delta C_p$ = 0.475 J/gK versus a value of 0.625 J/gK reported by Wunderlich and co-workers \cite{Wunderlich1983}, on the basis of several estimations meeting the standards of acceptable data. Furthermore, several studies exist on both polymeric \cite{WimbergerFriedl1996,Saiter2010} and non-polymeric glass formers \cite{Miller1997,Golovchak2012} showing the existence of intermediate plateaus before total recovery of equilibrium is achieved.

\begin{figure} [htb!]
\centering
\includegraphics[width=.35\textwidth, angle=270]{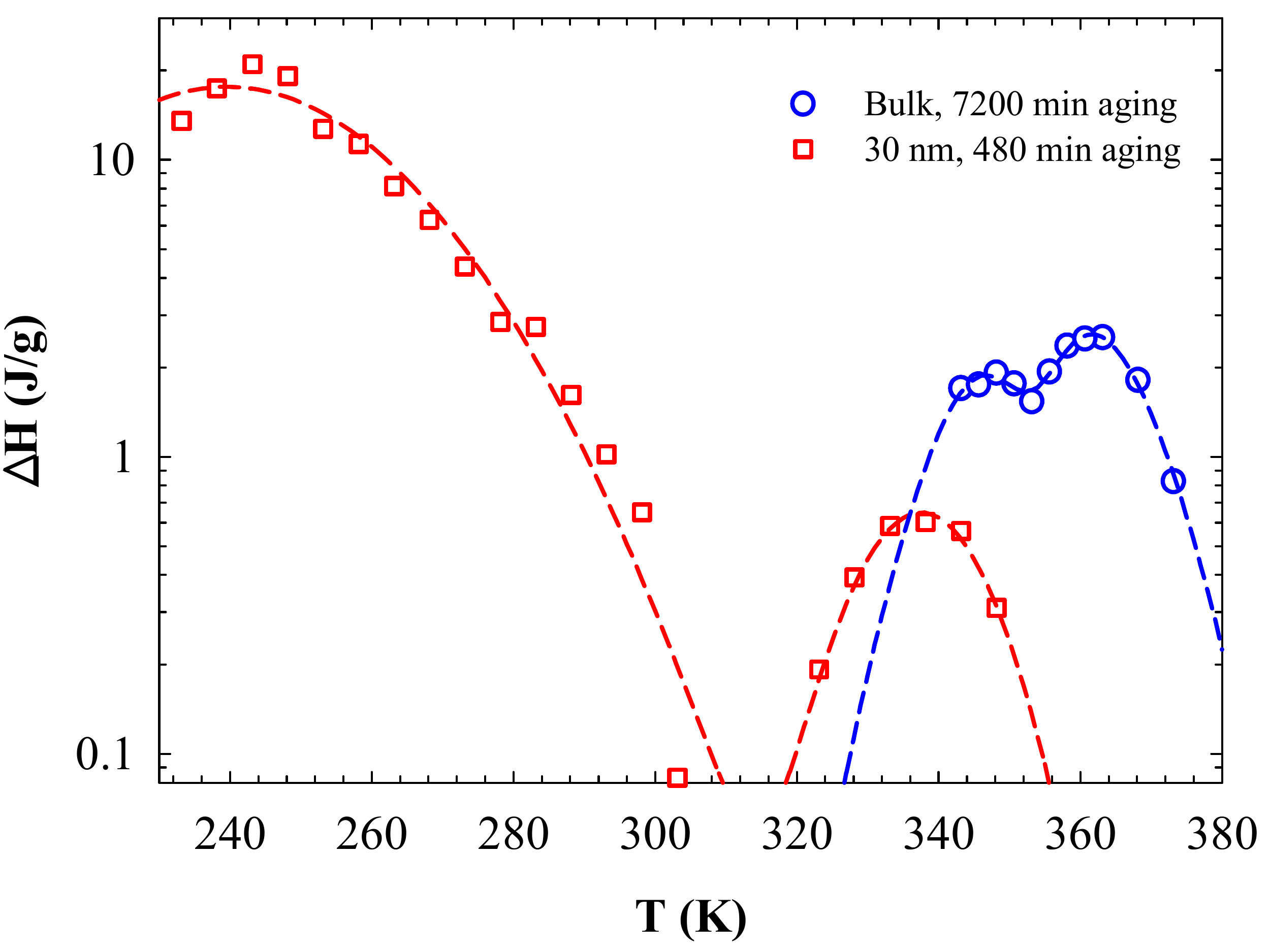}
\caption{Recovered enthalpy as a function of aging temperature for 30 nm thick PS films and bulk PS aged for 480 min and 5 days, respectively. Dashed lines are guides to the eye}
\label{Fig3SI}
\end{figure}

Here, the analogy between 30 nm thick PS and the corresponding bulk polymer for what concerns how the two mechanisms of equilibration act is presented showing unpublished results on isochronal enthalpy recovery experiments on bulk PS (5 days aging) at different temperatures. These data are plotted in Fig. \ref{Fig3SI} together with analogous data for 30 nm thick PS films aged for 480 min at different temperatures (already shown in the inset of Fig. 2 of the main manuscript). In both cases, the presence of two maxima in the plot of the recovered enthalpy as a function of temperature is indicative of the presence of two mechanisms of relaxation. However it is worth of remark that: i) the time scale of observation to achieve separation of the two time scales of equilibration is significantly larger for bulk PS than 30 nm films; ii) even with such a difference in the time scales of observation, the separation of the two maxima in the enthalpy is considerably more marked in the 30 nm thin films than in bulk PS. 

